\documentclass{ws-procs9x6}

\begin{document}

\title{Consistent discrete gravity solution of the problem of time: a model}

\author{Rodolfo Gambini, Rafael A. Porto}

\address{Instituto de F\'{\i}sica, Facultad de Ciencias\\
Universidad de la Rep\'ublica\\ Igu\'a esq. Mataojo, Montevideo,
CP 11400, Uruguay}

\author{Jorge Pullin}

\address{Department of Physics and Astronomy\\ Louisiana State
University\\ 202 Nicholson Hall, Baton Rouge LA 70803-4001, USA}

\maketitle
\abstracts{ The recently introduced consistent discrete lattice
formulation of canonical general relativity produces a discrete theory
that is constraint-free. This immediately allows to overcome several
of the traditional obstacles posed by the ``problem of time'' in
totally constrained systems and quantum gravity and cosmology. In
particular, one can implement the Page--Wootters relational
quantization.  This brief paper discusses this idea in the context of
a simple model system --the parameterized particle-- that is usually
considered one of the crucial tests for any proposal for solution to
the problem of time in quantum gravity.}

\section{Introduction}

There have always been problems to formulate general relativity on a lattice.
Even classically, if one simply discretizes Einstein's equations, the
resulting discrete equations are an inconsistent set of algebraic equations,
that is, they cannot be solved simultaneously. This is well known in
numerical relativity, where if one starts with a solution of the discrete
constraint equations initially, upon evolution with the discrete
evolution equations the resulting metric  and extrinsic curvature will
generically fail to solve the constraints.

A solution to this problem has recently been proposed\cite{DiGaPu,GaPuprl},
following the early work of T.D. Lee\cite{TDLee} in simple
mechanical systems. It consists in choosing the Lagrange multipliers
(in the case of the usual canonical formulations of general relativity
the lapse and the shift) so that the evolved quantities solve the
constraints. One has four equations for four unknowns so this is in
principle possible. This proposal has been explored in some detail in
the context of cosmologies\cite{GaPucosmo}, where the resulting
equations for the Lagrange multipliers are simple to solve. A canonical
formulation for these formulations has also been introduced\cite{DiGaPu}.

One of the attractive features of the resulting canonical theories is
that the constraints are gone, since they are solved for the Lagrange
multipliers.  The presence of the constraints is one of the major
roadblocks in the quantization of gravity. In particular most aspects
of the ``problem of time'' are related to the presence of the
constraints. The lack of constraints in the lattice formulation should
therefore allow to make considerable inroads into these problems.  We
will discuss this in general in a separate publication. In this short
paper we would like to exhibit how the construction works, in detail,
in a model system that many people consider to embody the germ of many
of the difficulties of the ``problem of time'' in quantum gravity: the
parameterized particle.

In particular we would like to exhibit how the definition of a
relational notion of time, as proposed in detail by Page and
Wootters\cite{PaWo}
(but with a long conceptual history going back to Leibniz, Mach,
Einstein, DeWitt, Hartle, Unruh, Barbour, Rovelli, Crane, Smolin and
many others) operates in detail in our proposal.

This paper is organized as follows. We start with a brief recap of
the proposal of Page and Wootters and the criticism it faced
(eloquently articulated by Kucha\v{r}\cite{Kuchar}). We then
briefly summarize in section 3 our consistent canonical
formulation of discrete mechanical systems. In the fourth section we apply
the construction to the parameterized particle. We end with a
summary and proposal for further implications of this formalism.

\section{The Page--Wootters construction}

The usual textbook formulation of quantum mechanics presupposes the
existence of an externally defined, classical quantity called ``$t$''.
The wavefunctions of the Schr\"odinger representation are functions
$\Psi(x,t)$. The variable $x$ has a very different role than $t$, it
is represented by an operator of which one can ---for instance---
compute expectation values. The time variable $t$, on the other hand,
remains always classical. This is clearly only an approximate concept
that requires the existence of a classical clock external to the
quantum system.  It is therefore not a very useful notion in the
context of closed systems where everything behaves quantum
mechanically, as for instance cosmology close to the Big Bang.

Page and Wootters\cite{PaWo} proposed an alternative to this. They
consider a system with variables $q_1, q_2,\ldots$. One then considers
wavefunctions $\Psi(q_1,q_2,\ldots)$ where all variables are now
quantum mechanical in nature and are represented by operators. One
then chooses one of the variables and calls it ``time'', for instance
$q_1=T$. Being a quantum variable, one now considers the calculation
of conditional probabilities in the sense of ``what is the probability
that the variable $q_k$ takes a value in a certain range $\Delta q_k$
near $q_k^0$ when ``time'' takes a value in a range $\Delta T$ near
$T^0$.'' This is how the Page--Wootters quantum mechanics is
constructed.

This egalitarian framework appears very attractive in the context of
quantum cosmology. However, there is a glitch. General relativity is a
constrained theory. In principle, one should consider classical
variables $q_1, \ldots$, etc. that are ``observables'' (that is, have
vanishing Poisson brackets with the constraints). Unfortunately since
one of the constraints is the generator of evolution, it means these
variables are ``perennials'' (as Kucha\v{r} usually calls them). This
implies that one cannot expect any of them to work as a clock. The
resulting quantum mechanics will have no evolution. Page and Wootters
tried to circumvent this by considering $q_1,\ldots$ to be
``kinematical'' variables (for instance the $g_{12}$ component of the
metric in some coordinates) that do not have vanishing Poisson
brackets with the constraints. In terms of these variables one would
expect to see ``evolution''. But here is where the presence of the
constraints generates problems. Kinematical variables do not have a
well defined action as quantum operators on states that are
annihilated by the constraints. Worse, such states are expected to
have a distributional nature as a subset of the full space of states.
This implies that they do not really admit a probabilistic
interpretation (see Kucha\v{r}\cite{Kuchar} for more details). This has a
concrete consequence.  When one uses the conditional probabilities of
Page and Wootters to compute a propagator, one gets the wrong result
that the system does not propagate (this can be seen in detail in the
example of the parameterized particle we discuss later\cite{Kuchar}).

Because of these problems and in spite of its initial naturalness and
appeal, the Page--Wootters proposal has been considered to fall short
of presenting a solution to the problem of time in quantum gravity.
As is made clear in the above discussion, it is the presence of the
Hamiltonian constraint in general relativity what really complicates
the application of this proposal. In the consistent discrete
canonical formulation of general relativity the constraints are not
present. Therefore it opens the possibility to revisit the Page and Wootters
proposal.

\section{Consistent discrete mechanics}

Since in this paper we will not really discuss general relativity but
only the example of the parameterized particle, it suffices to discuss
the consistent discrete approach in the context of mechanical systems
(in field theories additional subtleties appear, although the general
construction is similar\cite{DiGaPu}).

We start by considering a system with an action as in ordinary
classical mechanics $S=\int dt L(q,\dot{q})$. We then discretize
time, substituting the time derivative by
$\dot{q}=(q_{n+1}-q_n)/\Delta t$ and the integral in the action
becomes a sum. We write the action as $S=\sum_{i=1}^n
L(q_i,q_{i+1})$. One can work out the Lagrange equations. Then one
would usually introduce a Hamiltonian via a Legendre transform. We
would like to argue that this is not the most natural thing to do
in this context. After all, a Hamiltonian is an infinitesimal
generator of time evolution, and in a context where time is
discrete there is no natural meaning to an infinitesimal time
evolution generator. What is natural is to introduce a finite
generator. This is accomplished by a canonical transformation that
takes the system from $q_n$ to $q_{n+1}$. A generating function
$F$ of a type I canonical transformation (i.e. mapping from $q,p$
to $Q,P$ via $P=-\partial F(q,Q)/\partial Q$, $p=\partial F(q,Q)
/\partial q$), that implements the time evolution is simply given
by $F(q_i,q_{i+1})=-L(q_i,q_{i+1})\equiv -L(i,{i+1})$. The
evolution equations are therefore,
\begin{equation}
p_{n+1} = \frac{\partial L(q_n,q_{n+1})}{\partial q_{n+1}}
\qquad,\qquad p_n = - \frac{\partial L(q_n,q_{n+1})}{\partial q_n}
\,.\label{15}
\end{equation}
These equations define the canonical momenta $p$ and the evolution
for the canonical pair. The evolution preserves the Poisson bracket of
$q_i,p_i$. If we are a bit more explicit, writing out the usual form of
the Lagrangian for a constrained system,
\begin{equation}
L(n,n+1) = p_n (q_{n+1}-q_n) -H(q_n,p_n)-\lambda_{nB} \phi^B(q_n,p_n)
\end{equation}
where we assume we have $M$ constraints $\phi^B$ with $B=1\ldots M$
and $\lambda_{nB}$ are Lagrange multipliers. (We are slightly abusing the
notation; in order to have $M$ constraints one needs more than two
canonical variables, so one should assume that at each $n$ one has a
multicomponent vector for the $q$'s and $p$'s). If we work out the
equations of the canonical transformation we have,
\begin{eqnarray}
p_n-p_{n-1} &=& -{\partial H(q_n,p_n) \over \partial q_n} - \lambda_{nB}
{\partial \phi^B(q_n,p_n) \over \partial q_n}\label{43}\\
q_{n+1}-q_n &=&
{\partial H(q_n,p_n) \over \partial p_n}
+ \lambda_{nB} {\partial \phi^B(q_n,p_n) \over \partial p_n}\label{44}\\
\phi^B(q_n,p_n) &=&0.\label{45}
\end{eqnarray}

These equations look superficially like a straight discretization of the
usual Hamilton equations for constrained systems. But there is a significant
difference with the continuum case. The last equation (the constraints)
cannot hold simultaneously with the two first ones unless one chooses
specific ($p$ and $q$-dependent) values for the Lagrange multipliers.
With such a choice one is left with a free evolution given by the first
two equations with specific values of $\lambda$ and the constraints are
automatically satisfied.

Quantization of the system can be achieved by implementing the canonical
transformation as a unitary transformation quantum mechanically. For
instance, if one works in the Heisenberg picture, such a transformation
should implement equations (\ref{43},\ref{44}) as operator equations.
In the Schr\"odinger picture, wavefunctions are function of the discrete
parameter $n$.

It is deceiving to think that the above proposal for quantization will end
up being useful for systems like general relativity. Although one will
construct a quantum theory without conceptual obstructions (the only real
obstruction could be computational, i.e. solving the equations for the
Lagrange multipliers and then implementing the complicated resulting
classical evolution  equations operatorially via a unitary transformation)
the resulting theory would be difficult to interpret. The discrete parameter
$n$ has no direct physical meaning, it is just a computational entity
introduced in the discretization. Therefore it cannot be expected to be
correlated with any physically meaningful ``time'' in a generic situation.

In spite of these conceptual flaws, it should be emphasized that the
resulting quantum theory is well defined, with the wavefunctions having
a correct probabilistic interpretation. In this we see the first direct
positive consequence of not having constraints. The probabilistic
interpretation of the wavefunctions, which was problematic in
the Hartle--Hawking formulation (see Kucha\v{r}\cite{Kuchar} for a detailed
criticism) is suddenly not a problem here.

The fact that the most straightforward quantization does not appear to be
of physical interest, yet yields a good probabilistic interpretation
for the wavefunction is what makes the Page--Wootters approach quite
attractive. We have at hand the tools needed to compute the conditional
probabilities yet we will end up with a theory where the notion of time
is introduced physically and the parameter $n$ only plays a computationally
auxiliary role. Another way of putting it is that the discretization
technique naturally yields a concrete implementation of the
``mysterious time'' concept introduced by Unruh\cite{Unruhmoscow}.

Conceptually the proposal is therefore clear. The main point of this paper
is to show that the proposal actually works in detail in a simple
example that has however been quite problematic to tackle.

\section{The Page--Wootters quantization of the parameterized particle}

We cannot treat the free parameterized particle with our
technique. This is due to the fact that its Lagrangian (at least
written in the usual form) is simple enough  that the discrete evolution
equations are automatically consistent without applying our technique.
One could perform a change of variables to non-standard variables in
which the evolution equations could only be made consistent by determining
the Lagrange multipliers, but we will not pursue this route.

We prefer to consider a slightly more general (and in the end more
physically realistic) example: a non-relativistic particle in a constant
force field (for instance, a charged particle in a constant electric
field). The continuum Lagrangian can be written as,
\begin{equation}
S=\int  \left[p\dot{q}+p_0\dot{q}^0-N(p_0+{p^2\over
2m}+\alpha q)\right]d\tau,
\end{equation}
where the dot means derivative with respect to the parameter $\tau$,
$\alpha$ is the constant force on the particle due to the external
field, $N$ is the Lagrange multiplier. $q,p$ are the usual spatial
coordinate and momentum and $q^0,p_0$ are the temporal coordinate and
its conjugate momentum. The problem can be de-parameterized by
choosing $q^0=\tau$ and one recovers the ordinary mechanics of a
particle in a constant force field. If one chooses however $q=\tau$,
the resulting quantum theory is not consistent, yielding non-self-adjoint
operators\cite{parameter}.

We proceed to discretize the Lagrangian as we outlined in section 3,
\begin{equation}
L(n,n+1)=p^n(q_{n+1}-q_n)+p_0^n(q^0_{n+1}-q^0_n)-N_n(p_0^n+{p^2_n\over
2m}+\alpha q_n).
\end{equation}

As before, one can introduce a canonical transformation which implements
discrete time evolution. Its generator is $-L(n,n+1)$. The resulting
equations of the transformation are,
\begin{eqnarray}
P^{q^0}_n &=& P^{q^0}_{n+1}\label{ocho}\\ 
q^0_{n+1}-q^0_n &=& N_n
\label{nueve}\\
P^q_{n+1}-P^q_n&=& -\alpha N_n \label{diez}\\ 
q_{n+1}-q_n &=& N_n P^q_{n+1}/m
\label{once}\\
{P^q_{n+1}}^2/2m+P^{q^0}_{n+1}+\alpha q_n &=& 0.\label{constr}
\end{eqnarray}
In deriving these equations we have substituted the momenta canonically
conjugate to $p,p^0$ and $N$. The last equation would reproduce in
the continuum limit the usual constraint of the parameterized particle.
The Lagrange multiplier $N$ is determined
by solving the constraint (\ref{constr}),
\begin{equation}
N_n =\frac{m C_{n+1}}{\alpha P^q_{n+1}}, \label{lag}
\end{equation}
where,
\begin{equation}
C_{n+1}=
{{P^q_{n+1}}^2\over{2m}}+P^{q^0}_{n+1}+\alpha q_{n+1},
\end{equation}
is the discretization of the constraint of the continuum theory.

Which can be used to recast the equation of motion as an explicit canonical
transformation between $n+1 \to n$ as,

\begin{eqnarray}
q_{n}&=&q_{n+1}-\frac{C_{n+1}}{\alpha}\\
P^q_{n}&=&P^q_{n+1}+\frac{m C_{n+1}}{P^q_{n+1}}\\ 
q^0_{n}&=&
q^0_{n+1}-\frac{m C_{n+1}}{\alpha P^q_{n+1}}\\ 
P^{q^0}_n &=&P^{q^0}_{n+1}.
\end{eqnarray}

Notice that the discrete construction does not exist in the limit
$\alpha \to 0$. This prevent us from considering the free particle
case, as we argued above. The resulting theory has no constraint,
and therefore, distinctly from the continuum case, it has four
phase-space degrees of freedom (for a discussion of how to reconcile
a discrete theory with a given number of degrees of freedom as an
approximation to a continuum theory with a different number of
degrees of freedom see our recent paper\cite{GaPucosmo}).

We now proceed to quantize the system. We do it by constructing the
unitary transformation which reproduces in the Heisenberg language the
equations of motion. We consider wavefunctions $\Psi(q,q_0)$.

To begin with, we start by writing the following
matrix element,
\begin{equation}
<p'_0,p',n+1|p_0,q,n>=<p'_0,p',n|U|p_0,q,n>
\end{equation}
where $|p_0,q,n>$ , $<p'_0,p',n|$ are bases of the states at
instant $n$ labeled by the eigenvalues of ${\hat P}^{q^0}_n$,
${\hat q}_n$, and ${\hat P}^{q^0}_n$, ${\hat P}^{q}_n$,
respectively. Let us now to consider the operator version of
equation (\ref{constr}),
\begin{eqnarray}
&&<p',p'_0,n+1|{({\hat P}^q_{n+1})^2\over 2m}+{\hat P}^{q^0}_{n+1}+\alpha
{\hat q}_{n}|p,q,n>=\nonumber\\
&&({{p'}^2\over 2m}+p'_0+\alpha q)<p',p'_0,n+1|p,q,n>=0.\label{above}
\end{eqnarray}

This implies that
\begin{eqnarray}
<p',p'_0,n+1|p_0,q,n>&=&f(p'_0,p_0,q)\delta(p'+\sqrt{2mz})H(z)\\
&&+ g(p'_0,p_0,q)\delta(p'-\sqrt{2mz})H(z),\nonumber
\end{eqnarray}
with $z=-p_0-\alpha q$, $H(z)$ the step function, and $f(p',p,q)$ is a normalization factor to
be determined by the other set of equations in a similar fashion. Time
translation invariance of ${\hat P}^{q^0}$ is also naturally
implemented by replacing (\ref{ocho}) in (\ref{above}) which
immediately implies that it is proportional to $\delta(p'_0-p_0)$. On
a similar fashion, we implement the remaining equations
(\ref{nueve}-\ref{once}) where the Lagrange multiplier has been
substituted via equation (\ref{lag}) and a factor ordering has been
chosen. The final result for the evolution operator is,
\begin{eqnarray}
<p',p'_0,n|U|p_0,q,n>&=& A{1 \over z^{1/4}}
[\delta(p^0-p^0)\delta(p'+\sqrt{2mz})e^{-ipq}\\
&&-\delta(p^0-p^0)\delta(p'-\sqrt{2mz})e^{-ipq}]H(z).\nonumber
\end{eqnarray}

It can be seen that this operator is not strictly speaking unitary,
since $U^\dagger U =1$ but $UU^\dagger \neq 1$. It is what is
technically called an ``isometry''. A correct quantization can be
based on isometries\cite{cosmo2} , but we will not discuss this point
here. It should also be noticed that if one restricts the motion to
certain regions of configuration space before quantizing such that one
avoids the inversion point ($p=0$) in the motion, then the operator is
strictly unitary.

Although the quantization is complete, as we stressed before, its
physical interpretation is problematic since the ``evolution''
variable $n$ does not have any intrinsic meaning. The wavefunctions
however, have a correct probabilistic interpretations and therefore
can be used to compute the conditional probabilities of the Page--Wootters
relational approach.

Let us begin by introducing the conditional probability to obtain
$q=x$ given $q^0=t$ (strictly speaking we should phrase the
question in terms of intervals, since the operators have continuous
spectra, we omit them here to simplify the notation),
\begin{equation}
P(q=x|q^0=t)=\frac{\sum_{n=-\infty}^{\infty} |\Psi(x,t,n)|^2}
{\sum_{n=-\infty}^{\infty}\int_{-\infty}^{+\infty}dx|
\Psi(x,t,n)|^2}\label{prob1}
\end{equation}

Where we have used the standard formula, 
\begin{equation}
P_{\rm sim}(x,t) = P(t) P_{\rm cond}(x|t)
\end{equation}
where $P_{\rm sim}(x,t) =lim_{N \rightarrow \infty}{1\over{2N}}\sum_{n=-N}^N{ P_{\rm sim}(x,t)(n)}$, and $P_{\rm sim}(x,t)(n)$ is  the
probability of a $n$-{\it simultaneous} measurement of $q,q^0$.
The sum in $n$ is due to the lack of knowledge of the actual value
of $n$ when the measurement takes place. Generically, equation 
(\ref{prob1}) may face convergence difficulties. We will see that 
for the system in question, and for a judicious choice of time, no 
convergence problems appear.

Notice that the above definition of probability in principle allows other
choices of time variable, for instance,
\begin{equation}
P(q^0=t|q=x)=\frac{\sum_{n=-\infty}^{\infty} |\Psi(x,t,n)|^2}
{\sum_{n=-\infty}^{\infty}\int_{-\infty}^{+\infty}dt|\Psi(x,t,n)|^2}
\label{prob2}
\end{equation}
which would correspond to the computation of the ``time of arrival'',
a notoriously subtle problem (see for instance \cite{toa} and
references therein.)  This will require more detailed study, which we
will not attempt here, but it is interesting to notice that the
expression is well defined.

Let us now explicitly compute, for the system of interest, the
expression (\ref{prob1}). The calculation is not straightforward
unless certain approximations are made.  We need to compute the
operator $U$ explicitly (before we had just computed its matrix
elements). In order to do this, we write the operator as an
exponential of a ``Hamiltonian'' $U=e^{-iH}$ and therefore
$\hat{q}_{n}=e^{-iH} \hat{q}_{n+1} e^{iH}$. The explicit expression
for $H$ can be worked out as a power series in $C/{P^q}^2$. Here we
consider an approximation based on the leading term in the series. We
will also assume we are close to the continuum limit. The latter
occurs when the lapse goes to zero. By studying the solution to the
classical equations of motion (\ref{ocho}-\ref{once}) one can see that
the continuum limit is achieved in the limit in which $|P^q|>>|C|$
(far away from the turning point). In this limit, only the first term
of the power series contributes and the explicit expression for $H$
is,
\begin{equation}
H_{n+1} \sim {mC_{n+1}^2 \over 2\alpha P^q_{n+1}}.
\end{equation}

We can therefore quantum mechanically approximate the evolution
operator for $n$ steps $U(n) \equiv U^n$  by
$e^{-iHn}=\exp\left({-i{mC^2(p_0,p,q) \over 2\alpha p}n}\right)$,
with $C=p_0+h(q,p)$ the constraint, and in our case $h={p^2\over
2m}+\alpha q$.

We now would like to study states in which the variable $q^0$ behaves
approximately as a time variable, that is a wave-packet centered around
a given value with small dispersion and that does not spread significantly
upon evolution with respect to the parameter $n$.

We now expand the ``Hamiltonian'' as $H=\frac{{mp_0}^2}{2\alpha
p}+\frac{m h^2}{2\alpha p}+ \frac{p_0 m h}{\alpha p}$.  The first
term would produce dispersion in the variable $q^0$.  The introduction
of dispersion by the evolution in $n$ implies that for large values of
$n$ the system loses correlation between the time variable and $n$. In
a realistic quantum system, the clock variable will interact with
other degrees of freedom inducing decoherence and keeping small the
dispersion in the time variable. In our simple model, it is possible
to control the dispersion in $t$ without introducing additional
degrees of freedom by restricting the number N of discrete steps by
$N<<(\alpha p \sigma_t)/m$, where $\sigma_t$ is the initial standard
deviation of the time variable. This will allow us to use a sharply
peaked wave packet for the time variable in the whole range of values
of $n \in [-N,N]$. The second term is the final effective Hamiltonian
for the $q$-part of the system. The third term will generate the
evolution for the clock in the regime we are considering. Let us
examine more closely the denominator of (\ref{prob1}). Noticing that
$U^\dagger(n)|x,t,0>=|x,t,n>$ and similarly for the bra state, we have
that each term in the sum in the denominator is
\begin{eqnarray}
\int dx&& <\Psi|U^{\dagger}(n)|x,t,0><x,t,0|U(n)|\Psi>= \\&&\int dx
<\Psi|e^{iHn}|x,t,0><x,t,0|e^{-iHn}|\Psi>.\nonumber
\end{eqnarray}

Now let us assume an initially normalized wave packet of the form
$\Psi(x,t,n=0)=\phi_0(x)\psi_0(t)$, which would correspond to a weak
coupling between the clock and the particle. Ignoring
factor ordering corrections of order ${1\over p^2}$ in units 
$\hbar=c=1$ we get,
\begin{eqnarray}
\sum_n\int dx &&<\Psi|e^{ip_0{m h \over \alpha p}n}e^{i{m h^2 \over
2\alpha p}n} |x,t,0><x,t,0|e^{-ip_0{m h \over \alpha p}n}
e^{-i{m h^2\over 2\alpha p}n}|\Psi>=\nonumber\\
&&\sum_n<\psi_0|e^{ip_0<{m h \over \alpha p}>n}
|t,0><t,0|e^{-ip_0<{m h \over \alpha p}>n}|\psi_0>.
\end{eqnarray}
To derive this identity, we notice that $\exp(-i{m h^2\over 2\alpha
p})$ is an operator that only acts on $\phi_0(x)$ and therefore
can be transferred to the first inner product where it cancels
with $\exp(i{m h^2\over 2\alpha p})$. In the last step we have
neglected fluctuations on the expectation value of $\frac{m h}{\alpha
p}$ replacing it by $<\frac{m h}{\alpha p}>(0)$, the initial
expectation value, when integrating in $dx$. Thus, in the end, we get
 in the denominator the factor:

\begin{equation}
\sum_n |e^{-i{\hat p}_0<{m h\over\alpha p}>n}\psi_0(t)|^2=
\sum_n|\psi_0(t-<{m h\over\alpha p}>n)|^2.
\end{equation}

To appreciate better the meaning of this expression it is best to
replace the sum by an integral (since we are in the continuum
limit). In the continuum limit, it should hold that $<{m h\over\alpha p}>
<< T$, that is the elementary step in time should be much smaller than
the total period of validity of the relational approximation. Assuming
that the state is normalized, the integral yields a factor $<\frac{m
h}{\alpha p}>^{-1}$ for the denominator of the conditional
probability. Studying now the numerator in a similar fashion we end
with a term of the form,

\begin{eqnarray}
&\sum_n |e^{i{\hat p}_0<{m h\over\alpha p}>n}e^{i{m \hat h^2 \over
2\alpha{\hat p}n}}\phi_0(x)\psi_0(t)|^2=&\nonumber\\&\sum_n
|e^{i{m \hat h^2 \over 2\alpha{\hat p}}n} \phi_0(x)
\psi_0(t-<{m h\over\alpha p}>n)|^2,
\end{eqnarray}
where we have replaced the interaction term by its average value
assuming that $|<p>|>>\Delta p$, where $\Delta p$ is the initial
uncertainty of the wave packet in $p$. Now we may construct for the
initial condition a normalized wave function of the form
$\phi_0(t)=\delta_{\epsilon}(t)$ peaked around t=0.  Replacing again
the sum by an integral we cancel the denominator integrating the delta
function after replacing\footnote{Notice that in general we will not end with an exact
Schr\"odinger description, since the sum in $n$ may give more that one
factor if the dispersive effects and the decoherence by the
environment is included. This spreading effect of the clock
variable would imply information loss in the {\it time} evolution of
the wave packet. We have neglected this correction here in order to
compare with standard quantum mechanics. However, the information loss
is a natural consequence of the relational approach which may lead to
interesting new physics. We shall show this in a forthcoming
paper.} $n$ by ${t\over <{m h\over\alpha p}>}$. 
The final expression turns out to be:
\begin{equation}
P(q=x|q^0=t)=|e^{i{\hat{h}^2\over 2\alpha\hat{
p}}\times\frac{t}{<\frac{h}{\alpha p}>}}\phi_0(x)|^2\sim
|e^{i{\hat h^2 \over 2{\hat p}}\times<{h \over
p}>^{-1}t}\phi_0(x)|^2.
\end{equation}

Therefore we end with a Schr\"odinger picture that, as we shall see,
reproduces standard quantum mechanics. 
In order to see that the traditional
behavior is recovered, let us consider the Schr\"odinger picture
with $\hat{h}_{\rm eff}\equiv {\hat{h}^2 \over {2\hat p}}\times <{h
\over p}>^{-1}(0)$. Ignoring factor ordering terms, and also second
order terms in $1/p$, we
get, for the evolution of $<q>$ and $<p>$ the equations:

\begin{eqnarray}
{d \over dt}<q>={1 \over i}<[q,h_{eff}]>=<-i[q,h^2/2p]><{h \over p}>^{-1}(0)=\nonumber
\\{<{h \over p}>}^{-1}(0)[<(h^2/2p^2)>(t)+<h>(t)/m)]=\nonumber\\
<h/p>^{-1}(0)<h>(t)/m +1/2(<h^2/p^2>(t)/<h/p>(0)),\\ 
{d \over dt}<p>={1 \over i}<[p,h_{eff}]>=<-i [p,h^2/p]><{h \over p}>^{-1}(0)=\nonumber\\
-\alpha<{h\over p}>^{-1}(0)<{h\over p}>(t). 
\end{eqnarray}

The second equation reproduces the standard result ${d \over
dt}<p>=-\alpha$ for ${\alpha T \over <p>(0)} << 1 $, which ensures
that the expectation value $<h/p>(t)$does not depart  much from its
initial value during ``evolution". One can also recover the usual
result for the first equation ${d \over dt}<q>= <p>(0)/m$ provided
$<h>(0)/<p>(0)<<1$. This last requirement is a consequence of the two
previous conditions on $<p>$. One can show that it is possible to
assign numerical values to the different variables in such a way that
the complete set of conditions is satisfied for sufficiently long
periods of the $q^0$ time. One can also check that small uncertainty
states can be preserved during this period allowing to replace
operators for expectation values.

The correct quantum behavior is therefore extremely limited for this
system compared with the traditional Schr\"odinger quantization using
the Newtonian time. This is, however, expected since it is a direct
consequence of dealing with time as a quantum operator.  Furthermore,
for recovering the usual quantum behavior it is necessary to consider
the semiclassical properties of the clock including extra degrees of
freedom and the effects of the environment.  We have shown nonetheless
that the relational approach allows us for the first time to describe
nature in a completely quantum way. Notice as an interesting
consequence of the model that energy, in the form of $h$, is not
exactly conserved. In fact, $\frac{\Delta h}{h}\sim \frac{<\dot h>
T}{<h>} \sim \frac{\alpha}{<p>(0)}T << 1$ and therefore the relative
change in energy is small in the range of validity of the
approximation.

\section{Discussion}

We have shown that we can, through the use of the consistent
discretization approach, formulate a relational quantum mechanics for
the parameterized particle. As in the case of general relativity the
main ingredient is that the consistent discretization gets rid of the
constraint(s) of the system and therefore one can consistently
implement the relational description without the usual objections. The
example considered is therefore simple yet powerful. To our knowledge
there has never been a satisfactory relational quantum description in
the past for this system.

The relational description recovers usual quantum mechanics when the
discrete approximation approaches the continuum limit and when a clock
variable can be isolated that behaves sufficiently close to a classical
clock. The relational description does not require any of these
assumptions; it exists in much more general situations. In such situations
however, there is no sense in seeking a correspondence with usual
quantum mechanics.

The example we consider is important, but nevertheless limited. It
assumes the existence of a time variable whose simultaneity surfaces
are ``transverse'' to all the classical trajectories of the system. In
such situations it is known that the theory can be de-parameterized
and therefore one can solve immediately the problem of time (although
not through a relational quantum description). Even the same simple
example we have analyzed could be shown to exhibit non-transversality
by choosing the variable $q$. We will analyze this in a future
paper. We would also like to analyze a system where the coupling
between the clock and the system is less strong. This will require
more complexity, introducing extra degrees of freedom. A further
generalization we would like to study is systems where the constraint
does not have variables that appear linearly ---and therefore
de-parameterization is challenging---, as the Barbour-Bertotti model.
Unfortunately, the latter has a constraint that only depends on 
the momenta and therefore our discretization technique does not 
apply. A model where it could be tested is the $SL(2,R)$ model of
Montesinos, Rovelli and Thiemann\cite{MoRoTh} with two Hamiltonian
constraints.

Another problem that has received quite a bit of attention is the fact
that it could happen that the system is such that there are an
infinite number of crossings of the surface of simultaneity for a
given choice of time.  Indeed, for our approach this is a problem,
since in such a case the denominator of the conditional probability
blows up and therefore probabilities vanish. An example of this would
be to have a perfect oscillator as a clock. We see this as an
inescapable problem that, however, can be easily circumvented. A
``clock'' does not have to be associated with a single variable of the
problem. In general one will need a collection of variables to have a
good clock (just like one needs a wristwatch and a calendar to keep
time and date).  For instance two oscillators with frequencies whose
ratio is not a rational number would suffice. 

Another issue that has received some attention, for instance
emphasized by Unruh and Wald \cite{UW}, is that if one has a time
variable that is monotonous implies that the conjugate variable will
not be bounded below.  Since most Hamiltonians in physics are bounded
below, this appears as an obstruction. As we argued above, when we
chose $t=q$, we do not need a monotonous variable for our definitions
to exist (although their interpretation can be delicate). Furthermore,
a two-oscillator system with non-rational ratio of frequencies
behaves as a perfectly good monotonous clock despite having a Hamiltonian 
bounded from below. This is due to the fact that we may choose the
clock, in principle, as a function of any set of variables whose
Poisson algebra with $h$ would not have reason to be the canonical
one. In the example we worked out in detail we did not address this
problem since the Hamiltonian we considered is not bounded below.

A concern that might arise in the use of the discrete approach is 
that there are infinitely many ways of discretizing a continuum theory
and therefore the approach has a large degree of ambiguity. We see
this as inevitable. An analogy of interest is the case of the Feynman
path integral. There, even for simple systems, one defines the integral
via a discretization. Different discretizations are known to lead 
to the same theory, in different factor orderings. Our point of view
is therefore that the discretization ambiguities should be treated
as factor ordering ambiguities: they must be sorted by comparison
with experiment.

Summarizing, the consistent discrete approach to constrained systems, 
since it produces an approximate theory that is constraint-free, opens
the possibility of solving the problem of time through the construction
of a relational quantum mechanics. We have shown in detail how it works
for a simple example that captures some of the ingredients of the 
case of general relativity. To our knowledge this is the first time that
a correct relational quantum description has been constructed for this
system.

\section{Acknowledgments} 

We wish to thank Abhay Ashtekar, Julian
Barbour, Peter Haji\v{c}ek, Jim Hartle, Karel Kucha\v{r}, Carlo Rovelli,
Lee Smolin, Bill Unruh, for discussions and comments on the manuscript. 

This work was supported by grants NSF-PHY0090091, funds of the Horace
Hearne Jr.  Institute for Theoretical Physics, the Fulbright
Commission in Montevideo and PEDECIBA (Uruguay).

\end{document}